\documentclass[preprint]{aastex}
\shorttitle{Metal-Poor Globular Clusters}
\shortauthors{Strader, Brodie, \& Forbes}
\def\etal{{\it et al.}}
\input epsf

\begin{document}

\title{Metal-Poor Globular Clusters and Galaxy Formation}

\author{Jay Strader and Jean P. Brodie}
\affil{UCO/Lick Observatory, University of California, Santa Cruz, CA 95064}
\email{strader@ucolick.org, brodie@ucolick.org}
\and
\author{Duncan A. Forbes}
\affil{Centre for Astrophysics and Supercomputing, Swinburne University, 
Hawthorn VIC 3122, Australia}
\email{dforbes@astro.swin.edu.au}

\begin{abstract}

We demonstrate a significant ($> 5\sigma$) correlation between the mean color of \emph{metal-poor} globular cluster (GC) systems and parent galaxy luminosity. A Bayesian Markov
Chain Monte Carlo method is introduced to find the mean color, and is easily generalizable to quantify multimodality in other astronomical datasets. We derive a GC color--galaxy
luminosity relation of the form $Z \propto L^{0.15\pm0.03}$. When combined with evidence against a \emph{single} primordial GC metallicity--galaxy luminosity relation for
protogalactic fragments, the existence of such a correlation is evidence against both accretion and major merger scenarios as an explanation of the entire metal-poor GC systems
of luminous galaxies. However, our relation arises naturally in an \emph{in situ} picture of GC formation, and is consistent with the truncation of metal-poor GC formation by
reionization. A further implication is that the ages of metal-poor GCs in dwarf galaxies constrain the main epoch of galaxy formation in hierarchical models. If the ages of old
metal-poor GCs in Local Group dwarfs ($\ga 11$ Gyr) are typical of those in dwarfs elsewhere, then the bulk of galaxy assembly (at least in clusters and groups) must have
occurred at $z \ga 2.5$, contrary to the predictions of some structure formation models.
 
\end{abstract}


\section{Introduction}

A key development in the study of globular clusters (GCs) in external galaxies has been the discovery that most large galaxies have bimodal GC color
distributions (e.g., Zepf \& Ashman 1993; Forbes, Brodie \& Grillmair 1997; Gebhardt \& Kissler-Patig 1999; Kundu \& Whitmore 2001a; Larsen \etal~2001). These are usually
described as blue (metal-poor) and red (metal-rich) GC subpopulations, although additional substructure may be present. The red GC system properties appear
to be intimately tied to those of their parent galaxy, suggesting that the red GCs formed along with the bulk of the galaxy field stars (Forbes \etal~1997;
Forbes \& Forte 2001; Larsen \etal~2001). In both spirals and ellipticals, they are thought to be associated with the bulge/spheroid component (Forbes,
Brodie \& Larsen 2001).

The blue GCs are among the oldest and most metal-poor stellar systems observable. Therefore, they provide a probe of very early epochs of star formation in the universe.
Whether or not the properties of blue GCs correlate with the mass of their parent galaxy has been controversial (Forbes \etal~1997, C{\^ o}t{\' e} \etal~2000; Burgarella,
Kissler-Patig \& Veronique~2001; Forbes \& Forte~2001; Larsen \etal~2001, Lotz \etal~2003), and no clear demonstration of such a relation exists in the literature. However,
the issue is an important one in the context of GC and galaxy formation. If a correlation exists, it implies that the blue GCs, or at least a significant proportion of them,
``knew'' about the galaxy to which they would ultimately belong. This indicates that their formation was affected by the local conditions and that they may have already been
contained within the dark matter halo of their final galaxy. The detailed chemical and age structure within the blue GC systems of galaxies of various types and environments
would then offer one of the few observational constraints on the properties of the protogalactic clouds which combined to build the galaxies that we observe today. Such a
correlation would also rule out any formation mechanism whereby all metal-poor GCs form \emph{completely} independently of a host galaxy (e.g., Peebles \& Dicke 1968).

Our aim here is to consider galaxies over a large luminosity range, use only high-quality data, and analyze the photometry in a uniform manner. In this way we will reduce the
random and systematic errors that could disguise the existence of a blue GC--host galaxy correlation. In particular, we have added new data on the GC systems of dwarf and
low-luminosity elliptical galaxies, and utilized a Bayesian statistical method to find the peak of the blue GC color distribution.

\section{The Sample and Method}

We chose to explore a possible correlation between the mean $V-I$ color (i.e., the mode/peak of the Gaussian distribution) of metal-poor GC systems and the $M_{V}$ of
their host galaxies, since most GC system studies have been carried out in the $V$ and $I$ bands. When using only optical colors the age-metallicity degeneracy is
always a concern, but recent spectroscopic results suggest that, despite the presence of a small fraction of intermediate-age GCs in some early-type galaxies
(Goudfrooij \etal~2001; Larsen \etal~2003; Strader \etal~2003a), both metal-poor and metal-rich GC subpopulations appear to be very old ($\ga 10$ Gyr) within model
uncertainties (e.g., Larsen \etal~2002; Larsen \& Brodie~2002).

Our sources for massive early-type galaxies were Larsen \etal~(2001) and Kundu \& Whitmore (2001a, 2001b), who determined the peaks of the blue and red GC subpopulations by
fitting Gaussians as part of the KMM routine (Ashman, Bird, \& Zepf 1994). In addition, several nearby luminous spiral galaxies have enough blue GCs to have their mean
colors accurately determined. These include the Milky Way and M31 (Harris 1996, Barmby \etal~2000) as well as several Sculptor Group galaxies (Olsen \etal~2004). Our method
(see below) utilizes photometry for individual GCs, and we derive the peaks and errors ourselves rather than just using those reported in the literature. Therefore only
galaxies with high quality Hubble Space Telescope data (which has minimal contamination) and for which we had access to the photometry are included.

To probe the metal-poor GC systems of low-luminosity galaxies, we also included Local Group dwarf galaxies in our sample. These were
primarily taken from the compilation of Forbes \etal~(2000), though we have used new spectroscopic metallicities for old LMC GCs
(Beasley, Hoyle, \& Sharples 2002) whenever possible. The metallicities of Fornax GCs were taken from the study in Strader \etal~(2003b),
and we have added NGC 4147 to the metal-poor Sagittarius dwarf GCs (Bellazzini \etal~2003). The Local Group sample was supplemented with
the M81 dwarf~DDO78, which has one GC (Sharina, Sil'chenko, \& Burenkov 2003). Most of the dwarf GCs have spectroscopic and/or
color-magnitude diagram (CMD) based metallicities (presumably superior to those obtained from their $V-I$ colors), and these were
converted into $V-I$ colors using the Galactic relation of Barmby \etal~(2000). We included only genuinely \emph{old} GCs, excluding, for
example, intermediate-age GCs in the Magellanic Clouds. While further detections of GCs in dwarf galaxies outside the Local Group have
been claimed (e.g., in the M81 group; Karachentsev \etal~2000), we included only those whose identities have been confirmed by
spectroscopy. Finally, we note that since the majority of our sample galaxies are in groups or clusters, at present we can only claim to
be exploring the existence of a correlation in these environments.

For all galaxies with four or more GCs, we used Bayesian Markov Chain Monte Carlo (MCMC) methods, implemented in the package WinBUGS (Spiegelhalter \etal~2003), to
find the mean color of the blue GCs. See Gilks, Richardon, \& Spiegelhalter (1996) and Strader \& Brodie (2004) for more details on MCMC methods and for comparisons to
classical methods, e.g., maximum likelihood. For the luminous early-type galaxies, as well as the Milky Way and M31, we fit a homoscedastic (equal variance) Gaussian
mixture model to the $V-I$ colors to estimate the values of the blue and red peaks, as well as credible posterior intervals (the Bayesian equivalent of confidence
intervals) around those values. Indeed, the accurate estimation of these posterior intervals is one of the main advantages of using Bayesian method over the standard
least-squares framework. In the remainder of the galaxies, which had at most eighteen blue GCs, a single Gaussian was fit only to the metal-poor GCs (taken to be those
with [Fe/H] $< -1$). For galaxies with fewer than four GCs, the mean of the individual $V-I$ colors was used, with a fixed 0.05 mag (equivalent to $\sim 0.3$ dex)
1$\sigma$ error as a conservative estimate of the actual error.

Using WinBUGS, we fit a hierarchical linear model to the galaxy absolute magnitudes and mean GC colors. We made the \emph{a priori} choice of a linear model because such
models in ($M_V$, $V-I$) observational space reflect power-law models in luminosity/metallicity ($L$, $Z$) theoretical space. Our model has the following form:

\begin{eqnarray*}
        \overline{V-I}_{est_{i}} & = & \alpha + \beta M_{V_{i}} \\
        \overline{V-I}_{cosmic_{i}} & \sim & N(\overline{V-I}_{est_{i}}, \sigma^{2}_{cosmic}) \\
        \overline{V-I}_{i} & \sim & N(\overline{V-I}_{cosmic_{i}}, \sigma^{2}_{i})
\end{eqnarray*}

\noindent
\newline 
Diffuse (noninformative) priors were placed upon $\alpha$, $\beta$, and $\sigma^{2}_{cosmic}$. $\overline{V-I}_{est_{i}}$ is the initial estimated blue peak for the
$i^{th}$ galaxy given the priors on $\alpha$ and $\beta$. Cosmic variance and measurement errors are then hierarchically propagated to model the observed blue peak 
$\overline{V-I}_{i}$ for each galaxy.

\section{Results and Error Analysis}

\subsection{Results}

The resulting best fit weighted model is $V-I = (-0.0091\pm0.0018)M_{V} + (0.74\pm0.04)$, which is significant at the $5\sigma$ level. An unweighted model $V-I =
(-0.010\pm0.0015)M_{V} + (0.72\pm0.03)$ has a nearly identical slope at an even higher level of significance ($> 6\sigma$). This suggests no significant bias is introduced by our
calculated errors, and that the slope of the weighted model is not being defined by a small number of luminous galaxies with very accurately measured peaks. The nonparametric
Spearman rank correlation test confirms the correlation between $M_{V}$ and mean $V-I$ for blue GCs without assuming a specific model, finding a probability $p \le
4.8\times10^{-7}$ that there is no monotonic relationship between the two variables. We also separately fit weighted and unweighted models to only the spheroidal galaxies in our
sample (42 of the total 53 galaxies), since the younger stellar populations in disk galaxies may complicate our use of $M_{V}$ as a tracer of galaxy mass. The resulting models
are $V-I = (-0.0085\pm0.0019)M_{V} + (0.76\pm0.04)$ and $V-I = (-0.011\pm0.0015)M_{V} + (0.71\pm0.03)$, respectively, and very similar to those for the full sample.

The ``universal'' slope of $\sim -0.01$ found above is consistent at the $\sim 1\sigma$ level with the results of Larsen \etal~(2001), $-0.016\pm0.005$, and of Burgarella
\etal~(2001), $-0.009\pm0.002$. The bottom part of Figure 1 shows mean $V-I$ color for metal-poor GCs vs.~parent galaxy $M_{V}$.\footnote{In Figure 1, absolute magnitudes for
luminous early-type galaxies were taken from Prugniel \& Simien (1996), while Local Group galaxies came from Pritchet \& van den Bergh (1999). For Sculptor Group galaxies,
the distance moduli were from Karachentsev \etal~(2003), with photometry from Fitzgibbons (1990), Pierce \& Tully (1992), and de Vaucouleurs \etal~(1991). All Galactic
reddening corrections were from Schlegel, Finkbeiner, \& Davis (1998).}

While there are not enough disk galaxies to meaningfully derive a slope for these galaxies alone, an examination of Figure 1 suggests that they may have systematically
bluer metal-poor GC systems than spheroid-dominated galaxies: all eight disk galaxies with $M_{V} < -18$ fall below the best-fit model line for the full sample. The
apparent offset at $M_{V} = -20$ is $\sim 0.02$ mag, equivalent to just over $\sim 0.1$ dex in metallicity. This difference is likely partially due to our use of
$M_{V}$ as a tracer for galaxy mass, but an interpretation with a simple closed-box chemical evolution model also suggests that, at fixed luminosity, disk galaxies had
a larger gas fraction than spheroidal galaxies at the time their metal-poor GCs formed. This, in turn, implies that either (i) metal-poor GCs in disk galaxies are
older than their spheroidal counterparts, or (ii) an initial phase of chemical enrichment (to $\sim 0.03-0.05 Z_{\odot}$) proceeded more quickly in ellipticals than in
spirals.

In a recent simulation of GC formation in a hierarchical cosmology, Kravtsov \& Gnedin (2004) predict a relation close to $Z \propto L^{0.5}$ for GCs in giant
galaxies.  Though they do not distinguish between GC subpopulations, at the redshift that they stop their simulation ($z = 3.3$) all of their GCs are metal-poor
(with [Fe/H] $\la -1$), so a comparison to our relation is justified. For the purposes of this comparison, we convert the observed $V-I$ vs.~$M_{V}$
relation into $Z \propto L^{k}$ form using the Galactic color-metallicity relation $V-I = (0.156\pm0.015)\textrm{[Fe/H]} + (1.15\pm0.02)$ (Barmby \etal~2000). We
find $Z \propto L^{0.15\pm0.03}$, a relation that is substantially shallower than the Kravtsov \& Gnedin prediction.

\subsection{Error Analysis}

Could systematic statistical biases be responsible for our relation? First, we note that many of the possible sources of error apply \emph{only} to the low-luminosity galaxies in
our sample, e.g., the conversion from spectroscopic and CMD-based [Fe/H] to $V-I$ color. These possible biases can be addressed by using a homogeneous, unaffected subset of our
full sample. Taking only those galaxies that have $V-I$ HST data and that show visual bimodality, we find a best fit weighted model with a slope of --0.0098, nearly identical to
that obtained for the full sample. This shows that the leverage of the dwarf galaxies in our sample is not responsible for the correlation, in contrast
to the conclusions drawn by Burgarella \etal~(2001). Nor can \emph{any} systematic issue with the dwarf galaxies in our sample produce the observed relation.

One can then ask what systematic biases might affect our homogeneous subsample of luminous galaxies. Such biases could occur in two ways: (i) the detected correlation is a product
of selection effects of either the GCs in a given galaxy or the galaxies themselves, or (ii) the metal-poor GC peaks are being mismeasured in such a way as to produce our
observed relation. Consider point (i) first: the galaxy subsample, while by no means carefully chosen as a true volume or magnitude-limited sample, comprises a good
fraction of nearby luminous early-type galaxies, and it is reasonable to expect it to be generally representative of early-type galaxies in similar environments. Regarding the
selection of GCs in a given galaxy, our relation might arise if both galaxy luminosity and metal-poor GC metallicity were related through a third confounding variable.
Possibilities here include GC luminosity or galactocentric distance. Despite many studies which have searched for a GC metallicity-luminosity relation (e.g., van den Bergh 1996),
no such correlation has been observed in any galaxy to date. Furthermore, while GC systems as a whole have negative color gradients (i.e., become bluer at large radii), this
appears to be solely due to the differing radial distributions of the red and blue subpopulations, with little or no radial variation in the colors of the subpopulations
themselves (e.g., Lee, Kim, \& Geisler 1998; Harris, Harris, \& McLaughlin 1998). Thus, it does not appear likely that selection effects could have produced our relation.

Point (ii) above posits that our observed correlation is an artifact of the assumptions used to estimate the color distributions, namely, that the observed distributions can be
well-described as a superposition of two Gaussians with equal variance. Criticisms could be made of the assumptions of (a) bimodality, (b) Gaussian distributions, or (c)
homoscedasticity. Since galaxies in this subsample were selected to be \emph{visually} bimodal, as well as generally exhibiting strong statistical evidence against unimodality
(see Larsen \etal~2001), arguing against the assumption of bimodality requires instead that three or more subpopulations of GCs are present in the majority of the galaxies. No
convincing demonstrations of such multimodal systems have been made, even though some galaxies now have very deep HST data (e.g., M87; Jord{\' a}n \etal~2002). Assumption (b) can
be motived \emph{a priori} by the lognormal metallicity distributions predicted by a simple closed-box chemical evolution model, appealing to the Central Limit Theorem, or to the
dominance of Gaussian photometric errors. It could be argued that GC color distributions could have heavier tails than Gaussians, and be better fit by, e.g., Student's $t$
distributions (as the Galactic GC luminosity function appears to be; Secker 1992). As a test of this hypothesis, we fit the GC colors of M87 with a model very similar to that in
Sec.~2, except that we used $t$ distributions (instead of Gaussians) with the degrees of freedom as a free parameter. The posterior distributions of this parameter had most of
their mass at large values, suggesting that Gaussian fits are indeed appropriate (since a $t$ tends to a Gaussian as the degrees of freedom tend to infinity).

Regarding assumption (c) of homoscedasticity, due to the degeneracy between variance and other parameters in
mixture model fits, it is generally desirable assume a homoscedastic model unless there is \emph{a priori} evidence
against it (e.g., Ashman \etal~1994). For our sample, the blue and red GCs have similar luminosity functions (e.g.,
Larsen \etal~2001), so photometric errors could not be the source of unequal variances. Thus, any differences would
have to be intrinsic. The Galactic GC system, the only one which has metallicities measured for nearly all
clusters, appears to be homoscedastic within measurement errors (C{\^ o}t{\' e} 1999). In addition, any proposed
heteroscedasticity would need to be finely tuned, with an ever larger fraction of metal-rich GCs ``wrongly''
assigned to the metal-poor peak with increasing galaxy luminosity, to have any prospect of producing our relation.
As a simple numerical example, assume that all galaxies have a ``base'' mean blue GC color of $V-I$ = 0.90. Our
relation indicates that the most luminous galaxies in our sample have blue GC colors $V-I = 0.95$ and red GC colors
$V-I = 1.20$. Since the final peak is essentially a number-weighted average of the individual GC colors, and such
galaxies have similar numbers of blue and red clusters, an unreasonable $\ga 25\%$ of the red GCs would have to be
regularly misassigned to the blue peak to produce the colors we see for such galaxies (this is in addition to the
\emph{ad hoc} nature of the tuning noted above).

In sum, we find no evidence that systematic statistical biases could have led to the observed correlation between mean metal-poor GC color and parent galaxy luminosity.

\section{Implications and Discussion}

The classic Fall \& Rees (1988) classification of GC formation scenarios into primary, secondary, and tertiary has in the last decade given way to a new triad: accretion
(e.g., C{\^ o}t{\' e}, Marzke, \& West 1998), major merger (e.g., Ashman \& Zepf 1992), and \emph{in situ} (e.g., Forbes \etal~1997). The primary focus of each of these
formation scenarios is the GC systems of giant ellipticals (gEs). However, much of the following discussion applies also to the formation of less-luminous ellipticals (Es)
and disk galaxies.

C{\^ o}t{\' e} \etal~(1998, 2000, 2002) were able to produce bimodal metallicity distributions for the GC systems of luminous galaxies under the fundamental assumption that all
galaxies have one intrinsic population of GCs, which depends on galaxy luminosity. They used observations of a sample of dwarf and giant galaxies to find a ``primordial''
relation between GC metallicity and parent galaxy luminosity. By varying an assumed mass spectrum of protogalactic fragments, they were able to produce a wide variety of GC
metallicity distributions, some of which matched the bimodal form seen in most giant galaxies. It should be noted that they assume dwarf galaxies in the local universe (and
especially in the Local Group) are the surviving counterparts of these protogalactic fragments. If instead these gaseous fragments merged themselves out of existence at $\sim$
the epoch of blue GC formation, then the model is essentially equivalent to an \emph{in situ} scenario (see below).

The C{\^ o}t{\' e} \etal~(2002)~primordial mean metallicity-$M_V$ relation (see their Figure 1) is $V-I = -0.055M_{V} + 0.04$. This relation predicts that galaxies with $M_{V} =
-17.5$ will have one intrinsic population of GCs with $V-I = 1.00$. By contrast, the blue GC relation we derived above and the red GC relation from Larsen \etal~(2001) predict
peaks at $V-I = 0.90$ and $V-I = 1.09$, respectively. In Figure 1 we have overplotted red GC peaks for fifteen galaxies from Larsen \etal~(2001). To these we have added four
less-luminous galaxies (which are in our blue sample as well) for which we have remeasured the red peaks ourselves, using original data from Larsen \etal~and Kundu \& Whitmore
(2001a, 2001b). The dotted line is a new linear fit to the red peaks, weighted by the number of GCs in each peak. The dashed line is the C{\^o}t{\' e} \etal~relation. While the
observations are marginally consistent with the primordial relation at the blue and red ends of the luminosity range, in the intermediate-metallicity region between $M_{V} \sim
-17$ and $-18$ there is a clear, $\sim 0.1$ mag difference between the observed peaks and the primordial prediction. Thus, the C{\^ o}t{\' e} \etal~primordial relation for a
single intrinsic population of GCs appears to be less consistent with current observations than the existence of two separate (blue and red) GC metallicity--luminosity relations.

C{\^ o}t{\' e} \etal~(2002) showed that by using their primordial relation, a wide variety of GC color distributions could be produced. An examination of their Figure 1
suggests that, given appropriate input assumptions, the accretion model could reproduce the blue GC relation. However, the ability of the accretion model to successfully
reproduce our observed relation is critically dependent on the validity of the primordial relation. By contrast, a metal-poor GC--galaxy luminosity correlation arises
naturally under \emph{in situ} scenarios for GC formation, e.g., Harris \& Pudritz (1994) and Forbes \etal~(1997). In such scenarios, enrichment of the interstellar medium
is linked to the depth of the galactic potential, \emph{independent} of the details of GC formation in a particular scenario.

A blue GC--galaxy relation places limits on the degree to which an intrinsic population of metal-poor GCs in a
luminous galaxy has been affected by accretion and/or mergers, with substantial implications for galaxy formation
models (see below). A caveat is that the cosmic variance in our model (1$\sigma$ scatter of $\sim 0.02$ mag) does
not rule out the formation of the metal-poor GC systems of gEs like M49 and M87 from the accretion of several
slightly less luminous galaxies with redder-than-average blue GCs. However, our correlation implies that \emph{in
the mean} the blue GC systems of gEs could not have formed primarily from the accretion of less luminous galaxies,
especially $M_{V} > -18$ dwarfs, whose blue GCs have mean $V-I \la 0.90$. The paucity of metal-poor field stars in
the halos of luminous E galaxies (e.g., NGC 5128; Harris \& Harris 2000) is another argument against significant
accretion, since such stars would presumably accompany the accreted metal-poor GCs (Harris 2003). We are not
arguing that accretion does not take place; witness the existence of the Sagittarius dwarf (Ibata, Gilmore, \&
Irwin 1994), the recently discovered Canis Major galaxy (Martin \etal~2004; Forbes, Strader \& Brodie 2004), relic
stellar streams in the halo of M31 (Ibata \etal~2001), and more distant systems (e.g., Forbes \etal~2003). However,
we do suggest that it is not the primary mechanism by which the GC systems of most luminous galaxies are built.

We note that a major merger scenario for the formation of gEs is also difficult to reconcile with our results, since the mean colors of metal-poor GCs in typical Sb/Sc spirals
like the Milky Way and M31 are bluer than those of gEs (see Figure 1). The favored site for the formation of blue GCs in the major merger scenario is in pregalactic fragments
(Ashman \& Zepf 1992). Ashman (2003) states ``some variation in mean metallicity [of metal-poor GCs] would not rule out this option provided it did not correlate with properties
of the current parent galaxy''. As noted above for the accretion model, while cosmic variance would allow the occasional merger of two $L^{*}$ spirals with redder-than-average
blue GCs, the existence of a global correlation suggests that major mergers are unlikely to be the dominant mechanism for the creation of the GC systems of gEs. Independent
arguments in favor of this point have been made based upon the relative \emph{numbers} of metal-poor GCs in spirals and gEs (see Harris 2003 for a summary), though some recent
work (e.g., Rhode \& Zepf 2004)  suggests that these ``specific frequency'' arguments may only pose a problem for 
the blue GC systems of gEs.  We note that proponents of the merger scenario have themselves argued (see, e.g., the 
discussion in Ashman \& Zepf 1998) that major mergers cannot be the \emph{exclusive} mechanism for for forming gEs.

\emph{In situ} scenarios have been criticized with straw man comparisons to monolithic collapse models (e.g., Eggen, Lynden-Bell, \& Sandage 1962), though, for example, Forbes
\etal~(1997) argued that the initial collapse probably involved ``some chaotic merging of many small subunits'' as in the Galactic formation model of Searle \& Zinn (1978). This
idea was reflected in simulations of GC formation by Beasley \etal~(2002). One of the most significant problems with the \emph{in situ} scenario is the lack of a definitive
physical mechanism for the truncation of metal-poor GC formation. An initial starburst, accompanied by the formation of metal-poor GCs, could conceivably ionize and/or remove a
substantial mass of gas from inner regions of the protogalaxy, with the subsequent infall of enriched gas forming a population of metal-rich GCs and the bulk of the galaxy
starlight (Forbes \etal~1997). However, this feedback mechanism would be increasingly ineffective in more massive halos, with little suppression of star formation expected for
halos with virial velocities $V > 100$ km/s (Dekel \& Woo 2003). Thus, feedback is unlikely to be the source of the near-universal metallicity gap observed in the GC systems of
massive galaxies with $V \sim$ several hundred km/s. Cosmic reionization might also result in a rapid truncation of blue GC formation (e.g., Santos 2003), though some instead
argue for reionization as the trigger for metal-poor GC formation (Cen 2001). Among the advantages of the reionization hypothesis are its consistency with early structure
formation theories, as well as the prediction of the observed differences in metal-poor GC specific frequency between ellipticals and spirals. Since ellipticals (located in
high-density peaks) collapse earlier than spirals of similar mass, they have a larger fraction of their final mass assembled and available for GC formation before reionization.
By a similar argument, one might expect the metal-poor GC metallicity--galaxy luminosity relation to have a steeper slope in denser environments. A study comparing the blue GC
systems of field and cluster galaxies might then offer the intriguing possibility of elucidating the relative roles of galaxy mass and environment in shaping the initial phase of
local metal enrichment.

Our findings also have implications for popular hierarchical structure formation models (e.g., Kauffmann, White, \& Guiderdoni~1993; Cole \etal~2000; Somerville, Primack, \&
Faber 2001). For example, semi-analytic models generically predict a substantial fraction of mass assembly at $z \la 1$. However, as already noted, relatively luminous Es cannot
have accreted very many dwarf galaxies and their populations of GCs, since this would destroy our observed correlation. The ages of metal-poor GCs in dwarf galaxies place lower
limits on the possible epochs of significant accretion. We note that our results do not constrain (i) the accretion of galaxies before the formation of their GC systems, (ii) the
accretion of galaxies without GCs (which typically have $M_V > -13$), or (iii)  the formation of field galaxies. However, if the ages of old metal-poor GCs in Local Group dwarfs
($\ga 11$ Gyr; Johnson \etal~1999) are typical of those elsewhere (an untested assumption), then the bulk of galaxy assembly in groups and clusters must have occurred at $z \ga
2.5$. For more likely mean GC ages of $\sim 12$ Gyr, the primary epoch of mass assembly is pushed to $z \sim 4$ or higher.

\acknowledgements
WinBUGS code is available upon request from JS. We acknowledge support by the National Science Foundation through Grant AST-0206139 and a Graduate Research Fellowship (JS).
Graeme Smith, Soeren Larsen, Michael Beasley, Michael Pierce, and Michael West provided valuable comments on the manuscript. Suggestions from the referee Keith Ashman
significantly improved the paper. We thank Soeren Larsen, Arunuv Kundu, and Knut Olsen for providing us with their data.

\newpage

\epsfxsize=14cm
\epsfbox{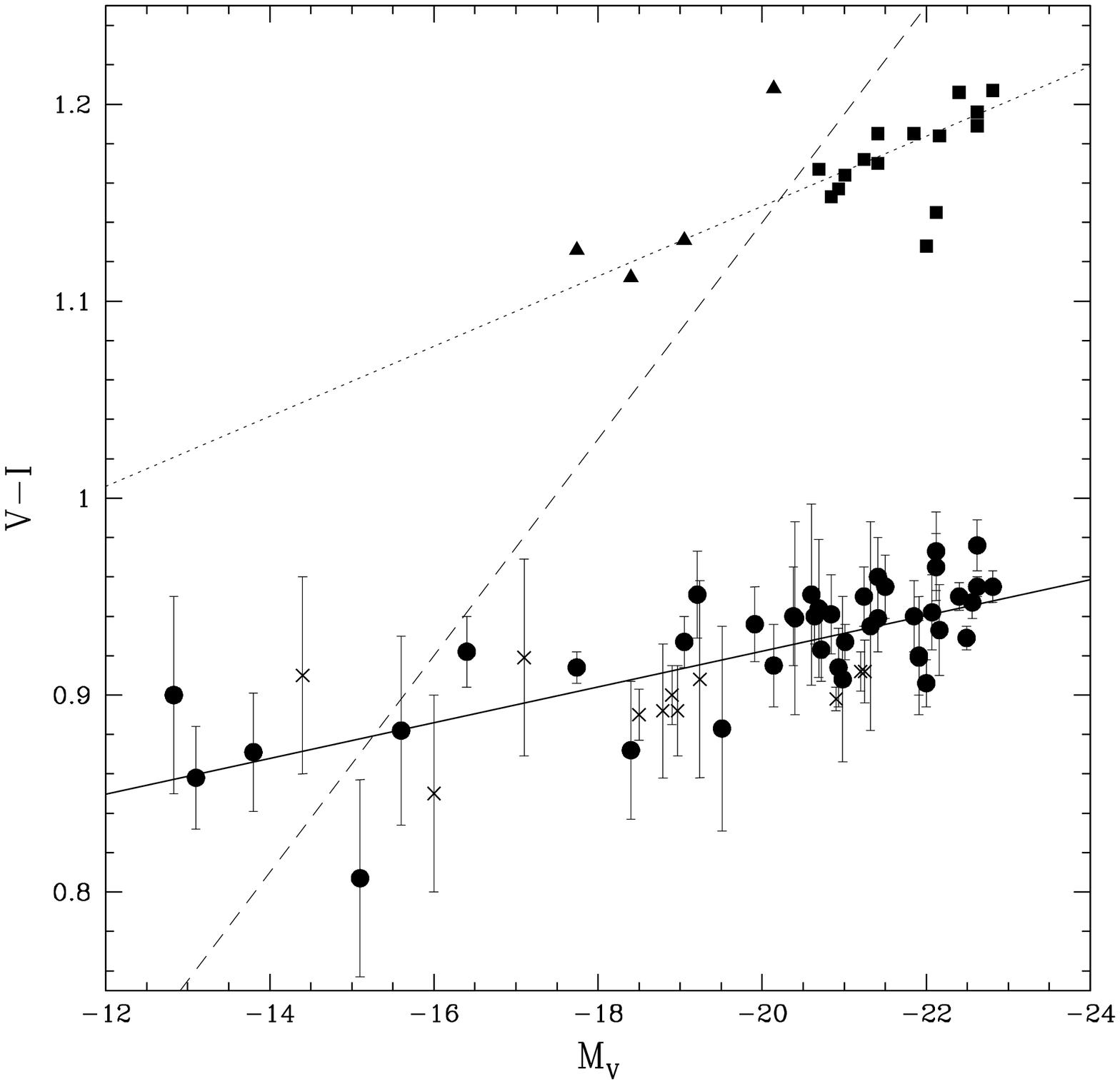}
\figcaption[strader.fig1.eps]{\label{fig:cor} Mean $V-I$ color of GC subpopulations vs.~luminosity ($M_{V}$) of their parent galaxies.  For
metal-poor GCs, the circles are spheroid-dominated systems, the crosses are disk-dominated, and the solid line is the best fit weighted
linear model for the full sample. The squares are metal-rich GC peaks from Larsen \etal~(2001), and the triangles are new metal-rich
peaks from this paper. The dashed line is the C{\^ o}t{\' e} \etal~(2002)~primordial mean GC metallicity-$M_V$ relation, and the dotted
line is a weighted linear fit to the red peaks.}

\end{document}